\documentclass[aps,final,notitlepage,oneside,onecolumn,nobibnotes,nofootinbib,
superscriptaddress,noshowpacs,centertags]{revtex4-1}
\usepackage{hyperref}
\usepackage{pxfonts}

\begin{document}
	
\title{Quadratic Gravity, double layers and non-conservative energy-momentum tensor}
\author{Victor A. Berezin}\thanks{e-mail: berezin@inr.ac.ru}
\affiliation{Institute for Nuclear Research of the Russian Academy of Sciences, Moscow, 117312 Russia}
\author{Vyacheslav I. Dokuchaev}\thanks{e-mail: dokuchaev@inr.ac.ru}
\affiliation{Institute for Nuclear Research of the Russian Academy of Sciences, Moscow, 117312 Russia}
\author{Yury N. Eroshenko}\thanks{e-mail: eroshenko@inr.ac.ru}
\affiliation{Institute for Nuclear Research of the Russian Academy of Sciences, Moscow, 117312 Russia}
\author{Aleksey L. Smirnov}\thanks{e-mail: smirnov@ms2.inr.ac.ru}
\affiliation{Institute for Nuclear Research of the Russian Academy of Sciences, Moscow, 117312 Russia}
	
\date{\today}

\begin{abstract}
In the present paper we investigate the conservative conditions in Quadratic Gravity. It is shown explicitly that the Bianchi identities lead to the conservative condition of the left-hand-side of the (gravitational) field equation. Therefore, the total energy-momentum tensor is conservative in the bulk (like in General Relativity). However, in Quadratic Gravity it is possible to have singular hupersurfaces separating the bulk regions with different behavior of the matter energy-momentum tensor or different vacua. They require special consideration. We derived the conservative conditions on such singular hypersurfaces and demonstrated the very possibility of the matter creation. In the remaining part of the paper we considered some applications illustrating  the obtained results.
\end{abstract}

\maketitle 
\maketitle 

\tableofcontents

\section{Introduction}
\label{intro}

Everybody knows that Albert Einstein imposing requirements to the relativistic theory of gravitation claimed that one of the most important among them is the conservative condition of the energy-momentum tensor --- the source of gravity. It is  this very feature of the (future) theory that allowed him to discover eventually the ``correct'' Einstein equations. It was more than one hundred years ago. Everybody believed at the time that the matter and geometry must be completely separated, and the matter can not be created.

The development of quantum theory revealed the fact that particles are created by strong fields from the vacuum state, and the vacuum itself is not empty but filled with zero fluctuations of all the fields. The gravitational field may cause the parametric resonance inducing the process of the matter creation. A.D.Sakharov even suggested \cite{Sakharov67} that the gravitational field does not exist as the fundamental one, but it is simply the tensions of the vacuum fluctuations of all other quantum fields.

In the early 70s of the last century three groups of theoreticians \cite{Zeld70,GribMamMost,ZeldPit71,ZeldStar,ParkerFulling,HuFullPar73,FullParHu74,FullPar74,LukashStar74,ZS77} started to investigate the processes of the quantum creation of particles by a scalar field on the cosmological model backgrounds. They found that the main role is played by the so-called conformal anomaly, which is a consequence of the renormalization procedure, necessary in quantum field theory.

The conformal anomaly ($=$ trace anomaly) can be incorporated into the action integral, where it consists of two parts, local and nonlocal ones. The local part enters the gravitational Lagrangian as the set of the counter-terms, and in the one-loop approximation equals to the sum of the terms, quadratic in Riemann curvature tensor and its contractions (Ricci tensor and curvature scalar). This is why we decided to consider just the Quadratic Gravity. Usually, the local part is interpreted as describing the vacuum polarization effects. The nonlocal part, then, associated with the particle creation process. While the local part explicitly inserted into the Lagrangian of the Quadratic Gravity, the nonlocal one is included, implicitly, into the matter Lagrangian and, consequently, into the energy-momentum tensor. If this picture is correct, it is natural to suppose that the local part gives rise to the divergentless gravitational field equations. Therefore, the corresponding part of the total energy-momentum tensor is conservative. We showed that this is indeed the case. Nevertheless, the transferring the geometry into the matter (particles) is possible due to that nonlocal term in the action integral mentioned above.

The above statement is valid provided the Riemann tensor and its contractions are well-defined, i.\,e., in the bulk. But, both in General Relativity and in the Quadratic Gravity there may appear singular hypersurfaces, where the curvature exhibit jumps and (Dirac's) delta-function behavior. In General Relativity this happens when the energy-momentum tensor is concentrated on this hypersurface, i.\,e., it has a term proportional to delta-function. In the Quadratic Gravity it is sufficient for the energy-momentum tensor to have a jump, describing the shock wave in the matter or even in the cosmological term (cosmological phase transition). These singular hypersurfaces are characterized by a three-dimensional metric tensor and three-dimensional extrinsic curvature tensor which describes the embedding of this three-dimensional hypersurface into the ambient four-dimensional space-time. 

The Einstein equations on the singular hypersurface were derived by W.Israel \cite{Israel66,Cruz67}. They relate the jumps in the extrinsic curvature tensor to the surface energy-momentum tensor of the thin shell formed by concentration of the matter fields on this hypersurface. In general, the surface energy-momentum tensor is four-dimensional. It can be, naturally, decomposed into the three-dimensional tensor, which is, actually, describes the energy content of the thin shell, the three-dimensional vector and the scalar. It appears that the last two, namely, vector and scalar parts are zero by virtue of the field equations.

In the case of Quadratic Gravity the corresponding equations were derived by J.M.M.Senovilla \cite{Senovilla13,Senovilla14,Senovilla15,Senovilla16,Senovilla17,Senovilla18}. They are essentially different from the Israel equations. First, the primeval equations contain not only the delta-function, but also its derivative. Thus, they describe not only the familiar thin shells, but also the so-called double layers. This is a rather new phenomenon, describing the gravitational shock waves. Second, after integration across the singular hypersurface there appear some ``arbitrary'' functions that should be determined by specifying the solutions in the bulk regions. Third, as it was mentioned and emphasized by J.M.M.Senovilla, the vector and scalar parts of the four-dimensional surface energy-momentum tensor are not necessarily zero, he gave them the names ``extrinsic pressure'' and ``extrinsic flow'', correspondingly.

In the present paper we derived the field equations on the singular hypersurface using quite different approach, namely, we did it using solely the least action principle. In the case of the Quadratic Gravity our method has an interesting feature: the derivative of the delta-function does not appear at all, and the delta-function itself appears only virtual. Nevertheless, we managed to extract the ``arbitrary'' functions. These functions enter only those equations that form the three-dimensional tensor, the three-dimensional surface energy-momentum tensor being the source. This means, that in order to determine the singular hypersurface itself for given solutions in the bulk we will need only the three-dimensional vector equation, as well as the scalar one. 

At last, we derived the conservative condition for the energy-momentum tensor on the singular hypersurface and proved that it is the vector and scalar parts of the surface energy-momentum tensor, that are responsible for the matter creation. One may say, that they are the ``remnants'' of the nonlocal part among the counter-terms leading to the conformal anomaly.

\section{Introduction to Quadratic Gravity}

The action integral for Quadratic Gravity, $S_2$, in its standard form reads as follows
\begin{equation} \label{S2}
	S_2=\int\!\mathcal{L}_2\sqrt{-g}\,d^4x,
\end{equation}
with the Lagrangian
\begin{equation} \label{L2}
	\mathcal{L}_2=\alpha_1R_{\mu\nu\lambda\sigma}R^{\mu\nu\lambda\sigma}+\alpha_2R_{\mu\nu}R^{\mu\nu}+\alpha_3R^2+\alpha_4R+\alpha_5\Lambda.
\end{equation}
Here $R^{\mu}_{\phantom{\mu}\nu\lambda\sigma}$ is the Riemann curvature tensor
\begin{equation} \label{Riemann}
	R^{\mu}_{\phantom{\mu}\nu\lambda\sigma}=\frac{\partial \Gamma^\mu_{\nu\sigma}}{\partial x^\lambda}-\frac{\partial \Gamma^\mu_{\nu\lambda}}{\partial x^\sigma}+\Gamma^\mu_{\varkappa\lambda}\Gamma^\varkappa_{\nu\sigma}-\Gamma^\mu_{\varkappa\sigma}\Gamma^\varkappa_{\nu\lambda},
\end{equation}
$R_{\nu\sigma}$ is the Ricci tensor
\begin{equation} \label{Ricci}
	R_{\nu\sigma}=R^{\lambda}_{\phantom{\mu}\nu\lambda\sigma},
\end{equation}
and $R$ is the curvature scalar
\begin{equation} \label{curvature}
	R=R^\lambda_{\lambda}
\end{equation}
and $\Lambda$ is the cosmological constant. We adopt the metric formalism, so the only gravitational dynamic variables are the components of the metric tensor $g_{\mu\nu}$ (and its inverse, $g^{\mu\nu}$: $g^{\mu\nu}g_{\nu\lambda}=\delta^\mu_\lambda$), defining the space-time interval, $ds$, 
\begin{equation} \label{ds2}
	ds^2=g_{\mu\nu}dx^\mu dx^\nu
\end{equation}
and the metric compatible symmetric connections ($=$ Christoffel symbols)
\begin{equation} \label{Christoffel}
	\Gamma^\lambda_{\mu\nu}=\frac{1}{2}g^{\lambda\varkappa}\left(g_{\varkappa\mu,\nu}+g_{\varkappa\nu,\mu}-g_{\mu\nu,\varkappa}\right),
\end{equation}
(comma (,) denotes a partial derivative).

The total action is the sum of the gravitational action and the the action of the matter fields $S_{\rm m}$, and the variation of the latter defines the matter energy-momentum tensor as follows
\begin{equation} \label{deltaSm}
	\delta S_{\rm m}=\frac{1}{2}\int 
	T_{\mu\nu}(\delta g^{\mu\nu})\sqrt{-g}\,d^4x =
	-\frac{1}{2}\int T^{\mu\nu}(\delta g_{\mu\nu})\sqrt{-g}\,d^4x.
\end{equation}

For our future purposes we will try to rewrite the gravitational action in the following way
\begin{equation} \label{L2b}
\mathcal{L}_2=\alpha\,C^2+\beta\,GB + \gamma R^2+\alpha_4R+\alpha_5\Lambda.
\end{equation}
Here $C_{\mu\nu\lambda\sigma}$ is the Weyl tensor 
\begin{equation} \label{Weyl}
	C_{\mu\nu\lambda\sigma}=R_{\mu\nu\lambda\sigma}+\frac{1}{2}(R_{\mu\sigma}g_{\nu\lambda}
	+R_{\nu\lambda}g_{\mu\sigma}-R_{\mu\lambda}g_{\nu\sigma}-R_{\nu\sigma}g_{\mu\lambda}) +\frac{1}{6}R(g_{\mu\lambda}g_{\mu\sigma}-g_{\mu\sigma}g_{\nu\lambda}),
\end{equation}
which is defined as the completely traceless part of the Riemann tensor, and its square $C_{\mu\nu\lambda\sigma}C^{\mu\nu\lambda\sigma}$ equals
\begin{equation} \label{C2}
	C^2=R_{\mu\nu\lambda\sigma}R^{\mu\nu\lambda\sigma} 
	-2R_{\mu\nu}R^{\mu\nu}+\frac{1}{3}R^2.
\end{equation}
Note, that all the coefficients in the last two formulas are calculated for the four-dimensional space-time.

Then goes for the famous Gauss--Bonnet term, $GB$,
\begin{equation} \label{GB}
	GB=R_{\mu\nu\lambda\sigma}R^{\mu\nu\lambda\sigma} 
	-4R_{\mu\nu}R^{\mu\nu}+R^2,
\end{equation}
which is pure topological in the four-dimensional space-time, i.\,e., after integration over some manifold $M$ it gives us its Euler characteristic $\chi$ (topological invariant) 
\begin{equation} \label{chi}
	\chi(M)=\frac{1}{32\pi^2}\int\limits_{M}GB\sqrt{-g}\,d^4x.
\end{equation}
Gauss--Bonnet term does not influence the field equations, provided the Riemann curvature tensor $R_{\mu\nu\lambda\sigma}$ is well defined. Evidently, special linear combination of $C^2$ and $GB$ can be used instead of any given combination of $\alpha_1$- and $\alpha_2$-terms in the standard form of the Lagrangian $\mathcal{L}_2$, and new coefficient $\alpha$ $\beta$ and $\gamma$ are
\begin{equation}\label{array}
	\left\{
	\begin{array}{l}	
		\alpha=2\alpha_1+\frac{1}{2}\alpha_2; \\ 
		\beta=-\alpha_1-\frac{1}{2}\alpha_2; \\
		\gamma=\alpha_3+\frac{1}{3}(\alpha_1+\alpha_2). \\
	\end{array}	 
	\right. 
\end{equation}

Let us now write the field equations. Surely, $\alpha_4$- and  $\alpha_5$- parts give Einstein equations (for  $\alpha_5=2\alpha_4=1/(8\pi G)$) with the cosmological term,
\begin{eqnarray} 
	&&\alpha_4 G_{\mu\nu}-\frac{1}{2}\alpha_5\Lambda g_{\mu\nu} =\frac{1}{2}T_{\mu\nu}(R),  \label{Einstein} \\
	&&G_{\mu\nu}=R_{\mu\nu}-\frac{1}{2}g_{\mu\nu}R,
	\label{Einstein2}
\end{eqnarray}
$G_{\mu\nu}$ is the Einstein tensor.

The $\alpha$-part in (\ref{L2b}) produces the so-called Bach equations,
\begin{eqnarray} \label{Bach}
B_{\mu\nu}&=&\frac{1}{2}T_{\mu\nu}(C^2), \\
B_{\mu\nu}&=&C_{\mu\lambda\nu\sigma}^{\phantom{0000};\sigma;\lambda}
+\frac{1}{2}R^{\lambda\sigma}C_{\mu\lambda\nu\sigma}, 
	\label{Bach2}
\end{eqnarray}
$B_{\mu\nu}$ is the Bach tensor, symmetric and traceless. Semicolon (;) denotes the covariant derivative with respect to the metric compatible connection, $\Gamma^\lambda_{\mu\nu}$, introduced above in (\ref{Christoffel}). 

The Gauss--Bonnet term --- no field equations whenever  $R^\mu_{\phantom{\mu}\nu\lambda\sigma}$ is well-posed.

At last, the $\gamma$-part provides us with the following field equations
\begin{eqnarray} 
	\mathcal{D}_{\mu\nu}&=&\frac{1}{4\gamma}T_{\mu\nu}(R^2), 
	\label{D} \\
	\mathcal{D}_{\mu\nu}&=&R_{;\mu\nu}-(R_{;\lambda;\kappa}g^{\lambda\kappa})g_{\mu\nu}-R(R_{\mu\nu}-\frac{1}{4}Rg_{\mu\nu}). 
	\label{D2}
\end{eqnarray}
On the right-hand-sides of field equations (\ref{D}) we put the corresponding partial energy-momentum tensor (notations are obvious).

\section{Introduction to double layers}

It was already mentioned that the field equations (presented in the previous Section~\ref{intro}) are valid only if the curvature tensor is well-defined, i.\,e., metric tensor is differentiable. But, one can easily imagine the physical situation when the energy-momentum tensor of the matter fields behaves differently in different regions of the space-time. In this case the energy-momentum tensor can be approximately described by the jumps (shock waves, matter-vacuum boundaries etc.) or by the Dirac $\delta$-function distribution (caustics, vacuum phase transitions etc). Here we suppose that there are no more singular terms (like $\delta'$, $\delta''$, \ldots) in the matter distribution. Therefore, we need to know the field equations not only in the bulk, but also on these transient surfaces which can be considered  mathematically as the junction conditions for the two different solutions in two different space-time regions.

Let the whole space-time is divided into two regions, $(+)$ and $(-)$, separated by some fixed $3$-dimensional (non-null) hypersurface $\Sigma_0$. Given the solutions in the bulk, i.\,e., in the $(\pm)$-regions, what should be the junction conditions for the metric tensor and its derivatives at $\Sigma_0$? It is well known that, by suitable coordinate transformation (of course, different in $(\pm)$-regions) it is always possible to make the metric tensor $g_{\mu\nu}$ continuous at any non-null hypersurface $\Sigma_0$. 

Let $n(x^\mu)=0$ be an equation of our hypersurface $\Sigma_0$ (of course, different in $(\pm)$-regions). Then, it is always possible to introduce in its vicinity the famous Gauss normal coordinate system, associated with $\Sigma_0$,
\begin{equation}\label{Gauus}
	ds^2=\epsilon\,dn^2+\gamma_{ij}dx^idx^j,
\end{equation}
where $n$ runs from $(-)$-region ($n<0$) to $(+)$-region ($n>0$) along the normal direction, $\epsilon=\pm1$, depending on whether the hypersurface $\Sigma_0$ is time-like ($\epsilon=-1$) or space-like ($\epsilon=1$). Our hypersurface is characterized by a $3$-dimensional metric $\gamma_{ij}$ (in what follows we adopt the $4$-dimensional signature $(+,-,-,-)$, the Greek indices take values $\{0,1,2,3\}$, while latin ones --- $\{0,2,3\}$ or $\{1,2,3\}$ and the extrinsic curvature tensor $K_{ij}$, which describes the embedding of a $3$-dimensional surface into a $4$-dimensional space-time. In the Gauss normal coordinates 
\begin{equation}
	K_{ij}= -\frac{1}{2}\gamma_{ij,n}.
\end{equation}

The matter energy-momentum tensor can be written now in the following way
\begin{equation} \label{Tmunu}
	T^{\mu\nu}=S^{\mu\nu}\delta(n)+T^{\mu\nu}(+)\theta(n)+T^{\mu\nu}(-)\theta(-n),
\end{equation}
where $\delta(n)$ is the Dirac $\delta$-function, and $\theta(n)$ is the Heaviside step-function. The tensor $S^{\mu\nu}$ is called the surface energy-momentum tensor. The behavior of the metric tensor derivatives at $\Sigma_0$ depends on the gravitational theory we are considering.

In General Relativity everything is rather simple. The field equations (\ref{Einstein}) are of the second order in derivatives of the metric tensor. With $S^{\mu\nu}\neq0$ and continuous $g_{\mu\nu}$ at $\Sigma_0$, the only way to compensate the $\delta$-function in the energy-momentum tensor is to demand the jumps of the extrinsic curvature tensor $K_{ij}$, 
\begin{equation}
	[K_{ij}]\neq0,
\end{equation}
where $[\phantom{i}]=(+)-(-)$. The junction conditions in this case are the Israel equations \cite{Israel66,Cruz67} that establish relations between $[K_{ij}]$ and $[S_{ij}]$.  If the surface energy-momentum tensor is nonzero, the hypersurface is singular, it has the special name ``thin shell''. In the case, when the energy-momentum tensor has only jumps (i.\,e., there is a shock wave in the matter distribution), then $\Sigma_0$ is non-singular, and the corresponding jumps in curvature mean that we are dealing with a shock gravitational wave.

In Quadratic Gravity the situation is more subtle. We are not allowed to have jumps in the first derivatives of the metric tensor (i.\,e., in the Christoffel symbols) --- otherwise in the Lagrangian (\ref{L2}) would appear $\delta^2$-function (in generic case), what ia absolutely forbidden in the conventional theory of distributions. Instead, we are forced to impose the so-called Lichnerowicz conditions 
\begin{equation}
	[g_{\mu\nu,\lambda}]=0,
\end{equation}
which in the Gauss normal coordinate system is transformed to 
\begin{equation}
	[K_{ij}]=0.
\end{equation}

Since in Quadratic Gravity the field equations are of the second order in derivatives of the curvatures (respectively, the fourth order in derivatives of the metric tensor), there are only two possibilities with $S^{\mu\nu}\neq0$. Either the Riemann curvature tensor is continuous at $\Sigma_0$, then its first derivatives undergo jumps, while the second derivative have a $\delta$-function behavior. In this case we are dealing with the thin shell, but the junction conditions ($=$ equations for the shell trajectory) will be quite different from that in General Relativity \cite{Frolov80}. Or the Riemann tensor undergoes a jump at $\Sigma_0$, then its first derivative have the $\delta$-function behavior, while the second derivative behaves like the $\delta'$ --- the derivative of the  $\delta$-function. And this what is called the ``double layer''. Note, that because of the jump in curvature it is, at the same time, the gravitational wave shock wave. It may be, or may not be, accompanied by thin shells.

Going further, we would like to show how the junction condition can be derived using the least action principle only. Let us start with General Relativity and Hilbert action
\begin{equation}
	S_{\rm H}=\alpha_4\!\int\!\!R\sqrt{-g}\,d^4x.
\end{equation}
Making the variation we get
\begin{eqnarray}
	\delta S_4&=&\alpha_4\int\!\Bigl\{(\delta R) 
	-\frac{1}{2}g_{\mu\nu}R(\delta g^{\mu\nu})\Bigr\}\sqrt{-g}\,d^4x \\
	&=&\alpha_4\int\!\Bigl\{g^{\mu\nu}(\delta R_{\mu\nu})+ 
	\frac{1}{2}(R_{\mu\nu}-g_{\mu\nu}R)(\delta g^{\mu\nu})\Bigr\}\sqrt{-g}\,d^4x.
\end{eqnarray}
For the variation of the Ricci tensor $\delta R_{\mu\nu}$ we will use the Palatini formula \cite{Palatini} \footnote{V.A.B. is greatly indebted to Prof. Friedrich Hehl, who told him about the author of this remarkable relation.}
\begin{equation}
	\delta R_{\mu\nu}= (\delta\Gamma^\lambda_{\mu\nu})_{;\lambda}- (\delta\Gamma^\lambda_{\mu\lambda})_{;\nu}
\end{equation}
(note that $\delta\Gamma$ is a vector). Thus,
\begin{equation}
	\delta S_4=\alpha_4\int\! \Bigl\{g^{\mu\nu}\Bigl((\delta\Gamma^\lambda_{\mu\nu})_{;\lambda} -\delta\Gamma^\lambda_{\mu\lambda})_{;\nu}\Bigr) -(R_{\mu\nu} -\frac{1}{2}g^{\mu\nu}R)(\delta g^{\mu\nu})\Bigr\}\sqrt{-g}\,d^4x.
\end{equation}
Both terms in the integrand contain $\delta$-function ($S^{\mu\nu}\neq0$), giving rise the contributions to the integral over the singular surface we are intersected in. We must carefully integrate them out. To do this, let us write the Christoffel symbols as  
\begin{equation}
	\Gamma^\lambda_{\mu\nu}=\Gamma^\lambda_{\mu\nu}(+)\theta(n)+\Gamma^\lambda_{\mu\nu}(-)\theta(-n).
\end{equation}
Then,
\begin{equation}
	\delta\Gamma^\lambda_{\mu\nu}=(\delta\Gamma^\lambda_{\mu\nu})(+)\theta(n)+(\delta\Gamma^\lambda_{\mu\nu})(-)\theta(-n)
\end{equation}
since we keep $\Sigma_0$ fixed, and
\begin{eqnarray}
	(\delta\Gamma^\lambda_{\mu\nu})_{;\lambda}&=& 
	[\delta\Gamma^\lambda_{\mu\nu})]\delta(n)n_{,\lambda}
	+(\delta\Gamma^\lambda_{\mu\nu})_{;\lambda}(\pm),
	\\
	(\delta\Gamma^\lambda_{\mu\lambda})_{;\nu}&=& 
	[\delta\Gamma^\lambda_{\mu\nu})]\delta(n)n_{,\lambda} 
	+(\delta\Gamma^\lambda_{\mu\lambda})_{;\nu}(\pm),
\end{eqnarray}
since $\theta(n)'=\delta(n)$ and $\theta(-n)'=-\delta(n)$. Analogously,
\begin{eqnarray}
	R_{\mu\nu}&=& 
	[\Gamma^\lambda_{\mu\nu}]\delta(n)n_{,\lambda}-
	[\Gamma^\lambda_{\mu\lambda}]\delta(n)n_{,\nu}+\ldots
	\\
	R&=& 
	g^{\alpha\beta}[\Gamma^\lambda_{\alpha\beta}]\delta(n)n_{,\lambda}-
	g^{\alpha\beta}[\Gamma^\lambda_{\alpha\lambda}]\delta(n)n_{,\beta}+\ldots
\end{eqnarray}

The simplest way to integrate across the singular hypersurface $\Sigma_0$ is to use the Gauss normal coordinate system. The result is straightforward
\begin{eqnarray}
	\delta S_4 &=& \alpha_4\int\limits_{\Sigma_0}\!\Bigl\{
	[\delta\Gamma^n_{\mu\nu}]g^{\mu\nu}
	-[\delta\Gamma^\lambda_{\mu\lambda}]g^{\mu n}\Bigr\}\sqrt{|\gamma|}\,d^3x
	\nonumber \\
	&+& \alpha_4\int\limits_{\Sigma_0}\!\Bigl\{
	[\delta\Gamma^n_{\mu\nu}](\delta g^{\mu\nu})
	-[\delta\Gamma^\lambda_{\mu\lambda}](\delta g^{\mu n}) 
	\nonumber \\
	&-&\frac{1}{2}(g^{\alpha\beta}[\Gamma^n_{\alpha\beta}]
	-g^{\alpha n}[\Gamma^\lambda_{\alpha\lambda}])g_{\mu\nu}
	(\delta g^{\mu\nu})
	\Bigr\}\sqrt{|\gamma|}\,d^3x 
	\nonumber \\
	&+& \alpha_4\int\limits_{(\pm)}\!\Bigl\{
	g^{\mu\nu}\bigl((\delta\Gamma^\lambda_{\mu\nu})_{;\lambda}
	-(\delta\Gamma^\lambda_{\mu\lambda})_{;\nu}\bigr)
	\Bigr\}\sqrt{-g}\,d^4x, 
	\label{deltaS4}
\end{eqnarray}
where $\gamma$ is the determinant of the $3$-dimensional metric $\gamma_{ij}$ on $\Sigma_0$. We have already omitted the terms in the volume integral, proportional to $\delta g^{\mu\nu}$, because they do not contribute to the surface integral anymore. Now, let us have a look at the remaining volume integral over $(\pm)$-regions only. It is easy to recognized that it consists of the linear combinations of full derivatives ($g^{\mu\nu}_{\phantom{\mu};\lambda}=0$, $l^\mu_{\phantom{;};\lambda}\sqrt{-g}=(l^\mu\sqrt{-g})_{\phantom{;},\lambda}$ for any vector $l^\mu$). By Stokes' theorem, it is converted into the surface integrals (here we need only the integral over $\Sigma_0$), namely 
\begin{eqnarray}
	&& \alpha_4\int\limits_{(\pm)}\!\Bigl\{
	(g^{\mu\nu}(\delta\Gamma^\lambda_{\mu\nu})\sqrt{-g})_{,\lambda}
	-g^{\mu\nu}(\delta\Gamma^\lambda_{\mu\lambda})\sqrt{-g})_{,\nu}
	\Bigr\}\sqrt{-g}\,d^4x \quad \rightarrow \nonumber \\
	&&-\, \alpha_4\int\limits_{\Sigma_0}\!\Bigl\{
	(g^{\mu\nu}[\delta\Gamma^\lambda_{\mu\nu}]-g^{\mu\lambda}[\delta\Gamma^\nu_{\mu\nu}]\Bigr\}\sqrt{-g}\,dS_\lambda,
\end{eqnarray}
where $dS^\lambda$ is the vector along the outward normal to  $\Sigma_0$. The sign in front of the integral is due to our definition of $[\phantom{i}]=(+)-(-)$. In Gauss normal coordinate system this becomes
\begin{equation}
	-\,\alpha_4\int\limits_{\Sigma_0}\!\Bigl\{
	g^{\mu\nu}[\delta\Gamma^n_{\mu\nu}]
	-g^{\mu n}[\delta\Gamma^\nu_{\mu\nu}]\Bigr\}\sqrt{|\gamma|}\,d^3x,
\end{equation}
i.\,e., exactly the same as the first line of equation (\ref{deltaS4}), but with the opposite sign! 

What do we have at the end of the day? Taking into account that 
\begin{equation}
	[\Gamma^n_{ij}]=\epsilon [K_{ij}], \quad [\Gamma^i_{nj}]=-[K^i_j],
\end{equation}
all others are zero, we have got ($K=Tr K_{ij}$)
\begin{equation} \label{Israel2}
	\alpha_4\epsilon\!\int\limits_{\Sigma_0}\!\Bigl\{
	[K_{ij}]-g_{ij}[K]\Bigr\}(\delta\gamma^{ij})\sqrt{|\gamma|}\,d^3x
	=-\frac{1}{2}\int\limits_{\Sigma_0}S_{\mu\nu}(\delta g^{\mu\nu})\sqrt{|\gamma|}\,d^3x,
\end{equation}
i.\,e., the Israel equations (for $\alpha_4=1/(16\pi G)$)
\begin{equation}
	\epsilon([K_{ij}]-g_{ij}[K])=8\pi G S_{ij}
\end{equation}
plus $S^{nn}=0$ and $S^{ni}=0$ due to the absence of $\delta g^{nn}$ and $\delta g^{ni}$ in the right-hand-side of equation (\ref{Israel2}).

Let us turn to the Quadratic Gravity. Here we are interested in deriving the junction conditions on the singular hypersurface $\Sigma_0$, using the least action principle only. The relevant calculations are very lengthy and cumbersome, so we will present here just main steps and will extensively make use of our experience in obtaining the Israel equations. For more details see \cite{Double19,Double20}.

So, our task is to find the contribution of the variation of the total action to the surface integral over some (fixed) singular hypersurface $\Sigma_0$. The gravitational action $S_2$ is defined in (\ref{S2}) with the Lagrangian $\mathcal{L}_2$ from equation (\ref{L2}). We already know that, beside the continuity of the metric tensor on $\Sigma_0$, what can be achieved by suitable coordinate transformation in $(\pm)$-regions, one has to impose, in addition, the so-called Lichnerowicz conditions, which, for given solutions in a bulk, serve, together with the junction conditions (still to be found) for determining this very singular hypersurface $\Sigma_0$. Thus, we demand that 
\begin{equation}
	[g_{\mu\nu}]=0, \quad [\Gamma^\lambda_{\mu\nu}]=0.
\end{equation}
Because of these Lichnerowicz conditions, there can be no $\delta$-functions in the Lagrangian $\mathcal{L}_2$, only jumps across $\Sigma_0$. For this very reason we can safely omit all the terms in $\delta S_2$ that proportional  to the variations of the metric tensor, $\delta g_{\mu\nu}$ ($\delta g^{\mu\nu}$), since they do not contribute to the surface integral, we are interested in here. Moreover, the $\alpha_4$-term will not contribute to the surface integral as well, what can be easily deduced from our preceding consideration. Thus we are left with 
\begin{equation}
	\delta S_2 \quad \rightarrow \quad
	2\int\!\!\Bigl\{\alpha_1R_\mu^{\phantom{\mu}\nu\lambda\sigma}(\delta 
	R^\mu_{\phantom{\mu}\nu\lambda\sigma})
	+\alpha_2R^{\mu\nu}(\delta R_{\mu\nu})
	+\alpha_3Rg^{\mu\nu}(\delta R_{\mu\nu})\Bigr\}\sqrt{-g}\,d^4x.
\end{equation}

We confine ourselves to showing some important details for the $\alpha_1$-patch. For the rest, $\alpha_2$- and  $\alpha_3$- ones we present only the final results.

So, we want to calculate the contribution to the surface integral over $\Sigma_0$ from the following volume integral
\begin{equation} \label{deltaS2alpha1}
	\delta S(\alpha_1)=2\alpha_1\!\int\! R_{\mu}^{\phantom{\mu}\nu\lambda\sigma}(\delta R^{\mu}_{\phantom{\mu}\nu\lambda\sigma})\sqrt{-g}\,d^4x.
\end{equation}
Again, we will make use of the Palatini formula, now for the Riemann curvature tensor,
\begin{equation}
	\delta R^{\mu}_{\phantom{\mu}\nu\lambda\sigma}=
	(\delta\Gamma^{\mu}_{\nu\sigma})_{;\lambda}
	-(\delta\Gamma^{\mu}_{\nu\lambda})_{;\sigma}
\end{equation}
Substituting this into the integrand in (\ref{deltaS2alpha1}) we get
\begin{equation}
	\delta  S(\alpha_1)=2\alpha_1\!\!\int\!\! R_{\mu}^{\phantom{\mu}\nu\lambda\sigma}
	\Bigl((\delta\Gamma^\mu_{\nu\sigma})_{;\lambda}
	-(\delta\Gamma^\mu_{\nu\lambda})_{;\sigma}
	\Bigr)\sqrt{-g}\,d^4x =4\alpha_1\!\!\int\!\! R_{\mu}^{\phantom{\mu}\nu\lambda\sigma}
	(\delta\Gamma^\mu_{\nu\sigma})_{;\lambda}\sqrt{-g}\,d^4x,
\end{equation}
where the symmetry property of the Riemann curvature tensor was taken into account. In the next step we extract the full derivative,
\begin{equation}
	\delta S(\alpha_1)=4\alpha_1\!\int \Bigl\{
	\left( R_{\mu}^{\phantom{\mu}\nu\lambda\sigma}(\delta\Gamma^\mu_{\nu\sigma})\right)_{;\lambda}-
	R_{\mu\phantom{\mu\mu\mu};\lambda}^{\phantom{\mu}\nu\lambda\sigma}(\delta\Gamma^\mu_{\nu\sigma})
	\Bigr\}\sqrt{-g}\,d^4x.
\end{equation}
Here, for the first time, we encountered with $\delta$-function. Indeed, 
\begin{equation}
	R_{\mu}^{\phantom{\mu}\nu\lambda\sigma}=R_{\mu}^{\phantom{\mu}\nu\lambda\sigma}(+)\theta(n)+R_{\mu}^{\phantom{\mu}\nu\lambda\sigma}(-)\theta(-n)
\end{equation}
and $[\delta\Gamma^\mu_{\nu\sigma}]=0$. Hence,
\begin{equation}
	\left( R_{\mu}^{\phantom{\mu}\nu\lambda\sigma}(\delta\Gamma^\mu_{\nu\sigma})\right)_{;\lambda}=
	[R_{\mu}^{\phantom{\mu}\nu\lambda\sigma}](\delta\Gamma^\mu_{\nu\sigma})\delta(n)n_{,\lambda}+\ldots
\end{equation}
and 
\begin{equation}
	\left( R_{\mu}^{\phantom{\mu}\nu\lambda\sigma}\right)_{;\lambda}
	(\delta\Gamma^\mu_{\nu\sigma})=
	[R_{\mu}^{\phantom{\mu}\nu\lambda\sigma}](\delta\Gamma^\mu_{\nu\sigma})\delta(n)n_{,\lambda}+\ldots
\end{equation}
($n(x^\mu)=0$ is an equation for $\Sigma_0$). Remarkably, they cancel each other in the integrand as it ought to be, because we started with the no $\delta$-function at all. Thus, we may forget about $\delta$-functions forever, and deal with the integrands over $(\pm)$-regions only, i.\,e., 
\begin{equation}
	\delta S(\alpha_1)=4\alpha_1\!\int\limits_{(\pm)} \Bigl\{
	\left( R_{\mu}^{\phantom{\mu}\nu\lambda\sigma}(\delta\Gamma^\mu_{\nu\sigma})\right)_{;\lambda}-
	R_{\mu\phantom{\mu\mu\mu};\lambda}^{\phantom{\mu}\nu\lambda\sigma}(\delta\Gamma^\mu_{\nu\sigma})
	\Bigr\}\sqrt{-g}\,d^4x.
\end{equation}
By making use of the Stokes' theorem we get 
\begin{equation}
	\delta S_{\rm gr}(\alpha_1)=-4\alpha_1\int\limits_{\Sigma_0}
	[R_{\mu}^{\phantom{\mu}\nu\lambda\sigma}](\delta\Gamma^\mu_{\nu\sigma})
	\sqrt{-g}dS_\lambda
	-4\alpha_1\int\limits_{(\pm)}
	R_{\mu\phantom{\mu\mu\mu};\lambda}^{\phantom{\mu}\nu\lambda\sigma} (\delta\Gamma^\mu_{\nu\sigma})
	\sqrt{-g}\,d^4x.
\end{equation}
(Note, again, the change of the sign in front of the surface integral.)

We do not intend to describe here the whole machinery, though it is by no means trivial, and present only the final result for the $\alpha_1$-patch:
\begin{eqnarray}
	\delta S_{\rm gr}(\alpha_1)&=& 4\alpha_1\!\int\limits_{\Sigma_0}\!
	\Bigl\{-2g^{il}g^{jp}[K_{lp,n}](\delta K_{ij})
	+\epsilon K^{lp}[K_{lp,n}](\delta g_{nn})
	\nonumber \\
	&&\!+2g^{il}g^{jp}[K_{lp,n|j}](\delta g_{in})
	+\!\Bigl(\!-g^{il}g^{jp}[K_{lp,nn}]-4g^{il}K^{jp}[K_{lp,n}]
	\nonumber \\
	&&\!+\!
	Kg^{il}g^{jp}[K_{lp,n}]\Bigr) (\delta \gamma_{ij}) \Bigr\}\sqrt{|\gamma|}\,d^3x,
\end{eqnarray} 
where the vertical line (|) denotes the $3$-dimensional covariant derivative.

In the same manner, i.\,e., without any details , we present the results for the $\alpha_2$- and $\alpha_3$-patches:
\begin{eqnarray}
	\label{deltaSgralpha2}
	\delta S_{\rm gr}(\alpha_2)&=&
	\alpha_2\!\int\limits_{\Sigma_0} \! \Bigl\{
	-2(g^{il}g^{jp}+g^{ij}g^{lp})[K_{lp,n} ] 
	(\delta K_{ij})  \nonumber \\
	&+&
	\!\!\epsilon\,(K^{lp}+Kg^{lp})[K_{lp,n}](\delta g_{nn}) +
	2(g^{il}g^{jp}+g^{ij}g^{lp})[K_{lp,n|j} ] (\delta g_{in}) 
	\nonumber \\
	&+& \!\!\Bigl(\,-\,(g^{il}g^{jp}+g^{ij}g^{lp})[
	K_{lp,nn}]+(-4g^{il}K^{jp}+2g^{lp}K^{ij} \nonumber \\
	&-&\!\!5g^{ij}K^{lp}+(g^{il}g^{jp}+g^{ij}g^{lp})K)[
	K_{lp,n}]\Bigr) (\delta\gamma_{ij}) \Bigr\} \sqrt{|\gamma|}\,d^3x,
\end{eqnarray}
\begin{eqnarray}
	\label{deltaSgralpha3fin}
	\delta S_{\rm gr}(\alpha_3)&=&\!\!4\alpha_3\!\int\limits_{\Sigma_0} \!
	\Bigl\{-2g^{ij}g^{lp}[K_{lp,n}] (\delta K_{ij}) + \epsilon Kg^{lp} 
	[K_{lp,n}] (\delta g_{nn})  
	\nonumber \\
	&+&\!\!2g^{ij}g^{lp}[K_{lp,n|j}]
	(\delta g_{in})+\Bigr(\!\!-g^{ij}g^{lp}[K_{lp,nn}]
	\nonumber \\
	&+&\!\!(-5g^{ij}K^{lp}+Kg^{lp}g^{ij}+K^{ij}g^{lp})[K_{lp,n}]
	\Bigr)(\delta\gamma_{ij})\Bigr\} \sqrt{|\gamma|}\,d^3x. 
\end{eqnarray}

In total,
\begin{eqnarray}
	\label{deltaSgralpha3fin}
	\delta S_{\rm gr}&=&\!\!\int\limits_{\Sigma_0}
	\Bigl\{ \Bigl\{-2\Bigl((4\alpha_1+\alpha_2)g^{il}g^{jp}+(\alpha_2
	+ 4\alpha_3)g^{ij}g^{lp}\Bigr)\Bigr\}[ K_{lp,n}] (\delta K_{ij}) 
	\nonumber \\
	&+&\!\epsilon\, \Bigl\{(4\alpha_1+\alpha_2)K^{lp} + (\alpha_2+4\alpha_3)Kg^{lp}
	\Bigr\}[K_{lp,n}](\delta g_{nn})  
	\nonumber \\
	&+&\!2\, \Bigl\{(4\alpha_1+\alpha_2)g^{il}g^{jp} 
	+(\alpha_2+4\alpha_3)g^{ij}g^{lp}\Bigr\}[K_{lp,n|j}](\delta g_{in})  
	\nonumber \\
	&+&\!\Bigl\{-\Bigl((4\alpha_1+\alpha_2)g^{il}g^{jp}
	+(\alpha_2+4\alpha_3)g^{ij}g^{lp}\Bigr)[K_{lp,nn}]  
	-4(4\alpha_1+\alpha_2)g^{il}K^{jp}[K_{lp,n}] 
	\nonumber \\
	&+&\! \Bigl((4\alpha_1+\alpha_2)g^{il}g^{jp} 
	+ (\alpha_2+4\alpha_3)g^{ij}g^{lp}\Bigr)K[K_{lp,n}]
	\nonumber \\
	&+&\!\Bigl((\alpha_2+4\alpha_3)(g^{lp}K^{ij}-5g^{ij}K^{lp})\Bigr)
	[K_{lp,n}]\Bigr\}(\delta\gamma_{ij}) \Bigr\} \sqrt{|\gamma|}\,d^3x. 
\end{eqnarray}

Let us analyze shortly the results obtained so far. 

First: since the variation of the total action on the singular hypersurface $\Sigma_0$ must be zero, then
\begin{equation}\label{deltaSgrsigma0}
	{\delta S_{\rm gr}}{\big|_{\Sigma_0}}=
	\frac{1}{2}\int\limits_{\Sigma_0} \! 
	S^{\mu\nu}(\delta g_{\mu\nu})\sqrt{|\gamma|}\,d^3x.
\end{equation}

Second: note, that the coefficients $\alpha_1$, $\alpha_2$ and  $\alpha_3$ from the Quadratic Gravity Lagrangian (\ref{L2}) enter
$\delta S_{\rm gr}(\Sigma_0)$ only in two combinations, $(4\alpha_1+\alpha_2)$ and $(\alpha_2+4\alpha_3)$. If both of them are zero, then $S^{\mu\nu}=0$. But, this is just the case of the pure Gauss-Bonnet term. Thus, if the Riemann curvature tensor does not exhibit the $\delta$-function behavior at $\Sigma_0$ (i.\,e., either it is continuous there, or undergoes a jump), then adding the Gauss-Bonnet term to the curvature scalar in the Hilbert action does not produces neither double layer, nor additional thin shells. Note also, that in this case the existence of the $\delta$-function in the curvature is possible (and, therefore, the Lichnerowicz conditions are not obligatory), because the $\delta^2$-terms do not appear in the corresponding Lagrangian.

Third: consider the case when the curvature is continuous at $\Sigma_0$, i.\,e., $[K_{ij,n}]=0$, and no double layer exists at all. Then, 
\begin{equation}\label{nojumps}
	- \!\int\limits_{\Sigma_0}\Bigl\{ 
	(4\alpha_1+\alpha_2)g^{il}g^{jp}+(\alpha_2 
	+4\alpha_3)g^{ij}g^{lp}\Bigr\}[K_{lp,nn}]
	(\delta\gamma_{ij}) \sqrt{|\gamma|}\,d^3x 
	=\frac{1}{2}\int\limits_{\Sigma_0} \! S^{\mu\nu}(\delta g_{\mu\nu})\sqrt{|\gamma|}\,d^3x.
\end{equation}
and
\begin{equation}\label{array2}
	\left\{
	\begin{array}{l}	
		\!-\,\Bigl\{(4\alpha_1+\alpha_2)g^{il}g^{jp}
		+(\alpha_2 +4\alpha_3)g^{ij}g^{lp}\Bigr\}
		[K_{lp,nn}] = \frac{1}{2}S^{ij}; \\ 
		\!S^{nn}=0, \; S^{ni}=0; \\
		\![ K_{ij}]=0, \;[ K_{ij,n}]=0.
	\end{array}	 \right.
\end{equation}
This is the analogue of the Israel equations for the thin shells.

At last, let us come to the generic case, when there is a jump in the curvature at the singular hypersurface, and, thus, the double layer is produced. At once, we encounter the problem. In the integrand there exist variations of the extrinsic curvature tensor $\delta K_{ij}$, which are not the variations of the dynamical variables and, at the same time, cannot be removed. In General Relativity these variations were canceled by contributions from the $\delta$-function terms in the Lagrangian, but now we completely lack such a possibility. What to do? The solution of this puzzle lies in recognizing, that $\delta K_{ij}$ are not the independent variations, they depend on $\delta\gamma_{ij}$, simply because $\delta K_{ij}=(1/2)(\delta\gamma_{ij,n})|_{\Sigma_0}$. But, in a sense, the relation between them is arbitrary, since the equations in the bulk, i.\,e., in $(\pm)$-regions, are of the fourth order in derivatives of the metric tensor, and they are not uniquely defined by $g_{\mu\nu}$ and $g_{\mu\nu,\lambda}$ at some Cauchy hypersurface. Thus we are forced to write down the following 
\begin{equation}\label{Bij}
	\delta K_{i'j'}=B_{i'j'}^{\phantom{ijj}ij}(\delta\gamma_{ij}).
\end{equation}
The appearance of the arbitrary function are not completely surprising. This is just a reminiscent of the $\delta'$-functions in the field equations, and, thus it is a marker of the double layer. In the next Section~\ref{Conservative} we will demonstrate, how it works. Actually, 
$B_{i'j'}^{\phantom{ijj}ij}$ are not completely arbitrary, they depend on our choice of the solutions in the bulk and should be found when solving the junction equations.

Now we are ready to write down the equations for the double layer in Quadratic Gravity \cite{Double19,Double20}:
\begin{eqnarray}
	\label{QG1}
	&&\epsilon\,\Bigl\{(4\alpha_1+\alpha_2)K^{lp}
	+(\alpha_2+4\alpha_3)Kg^{lp}\Bigr\} [K_{lp,n}]
	= \frac{1}{2}S^{nn},  \\
	\label{QG2}
	&&2\,\Bigl\{(4\alpha_1+\alpha_2)g^{il}g^{jp}
	+(\alpha_2+4\alpha_3)g^{ij}g^{lp}\Bigr\} K_{lp,n|j}]
	= \frac{1}{2}S^{in},  \\ 
	&&\Bigl\{\Bigl(-2(4\alpha_1+\alpha_2)g^{i'l}g^{j'p} + (\alpha_2+4\alpha_3)g^{i'j'}g^{lp} \Bigr)[ K_{lp,n}]
	B_{i'j'}^{\phantom{ijj}ij} 
	\nonumber \\
	\label{QG3}
	&&+\Bigl\{-\Bigl((4\alpha_1+\alpha_2)g^{il}g^{jp}
	+(\alpha_2+4\alpha_3)g^{ij}g^{lp}\Bigr)[K_{lp,nn}]  
	\nonumber \\
	&&-4(4\alpha_1+\alpha_2)g^{il}K^{jp}[K_{lp,n}] 
	\nonumber \\
	&&+ 
	\Bigl((4\alpha_1+\alpha_2)g^{il}g^{jp} 
	+ (\alpha_2+4\alpha_3)g^{ij}g^{lp}\Bigr)K[K_{lp,n}]
	\nonumber \\
	&&+\Bigl((\alpha_2+4\alpha_3)(g^{lp}K^{ij}-5g^{ij}K^{lp})\Bigr)
	[K_{lp,n}]\Bigr\}(\delta\gamma_{ij}) \Bigr\} = \frac{1}{2}S^{ij}.
\end{eqnarray}

J. M. M. Senovilla \cite{Senovilla13,Senovilla14,Senovilla15,Senovilla16,Senovilla17,Senovilla18}) was the first who discovered and emphasized the fact, that  $S^{nn}$ and $S^{ni}$ are not necessary zero. The structure of equations (\ref{QG1})--(\ref{QG3}) are rather curious. The $(nn)$ and $(ni)$ equations serve for determining the singular hypersurface $\Sigma_0$ itself, while the $(ij)$ equations serve for calculating the ``arbitrary'' functions $B_{i'j'}^{\phantom{ijj}ij}$ only.

\section{Conservative condition}
\label{Conservative}

More than one hundred years ago, Albert Einstein, trying to construct a relativistic theory of gravity, claimed that one of the requirement to the future theory must be the conservative condition for the energy-momentum tensor for the matter fields. He was, evidently, encouraged by the fact that in the case of Special Relativity the energy-momentum tensor $T^{\mu\nu}$ for the isolated  (conservative) systems automatically obeys the following conservation equation (in Minkowski coordinates)
\begin{equation}\label{Minkowski}
	T^{\mu\nu}_{\phantom{\mu\nu},\nu}=0,
\end{equation}
and, also, that Maxwell equations for the electrodynamics incorporates the electric charge conservation law (in Minkowski coordinates)
\begin{equation}\label{current}
	J^\nu_{\phantom{\mu},\nu}=0,
\end{equation}
where $J^\nu$ is the electric current $4$-vector. In Special Relativity the conservation law (\ref{Minkowski}) follows from the Noether's theorem due to the homogeneity of the flat space-time. When using curvilinear coordinates (but still in the Minkowski space-time), the conservative condition reads 
\begin{equation}\label{Minkowski}
	T^{\mu\nu}_{\phantom{\mu\nu};\nu}=0,
\end{equation}
where a semicolon denotes the covariant derivative with respect to the metric compatible connections (Christoffel symbols).

It is this condition that automatically holds in General Relativity due to the conservative condition for the Einstein tensor
\begin{equation}
	G^\nu_{\mu;\nu}=R^\nu_{\mu;\nu}-\frac{1}{2}R_{,\mu}=0,
\end{equation}
that follows from the Bianchi identities for the Riemann curvature tensor.
\begin{equation}
	R^\mu_{\phantom{\mu}\nu\lambda\sigma;\varkappa}
	+R^\mu_{\phantom{\mu}\nu\varkappa\lambda;\sigma}
	+R^\mu_{\phantom{\mu}\nu\sigma\varkappa;\lambda}=0.
\end{equation}
Indeed, the contraction in indices $\mu$ and $\varkappa$ gives us 
\begin{equation}
	R^\mu_{\phantom{\mu}\nu\lambda\sigma;\mu}
	+R_{\nu\lambda;\sigma}+R_{\nu\sigma;\lambda}=0,
\end{equation}
and the subsequent contraction in indices  $\nu$ and $\sigma$ leads to the famous relation 
\begin{equation}
	R^\nu_{\lambda;\nu}=\frac{1}{2}R_{;\lambda}.
\end{equation}
Thus, the corresponding part of the total energy-momentum tensor, i.\,e., $T^\mu_\nu(R)$ is conservative,
\begin{equation}
	T^{\mu\nu}(R)_{;\nu}=0.
\end{equation}

Let us turn to the next part, namely, to $T^{\mu\nu}(C^2)$. To check its conservative condition, we need to calculate the full covariant derivative of the Bach tensor, $B^{\mu\nu}_{\phantom{\mu};\nu}$. We do not intend to exhibit all the details, show only the main steps. Let us demonstrate, first, that the Bach tensor (\ref{Bach2}) is symmetric, $B^{\mu\nu}=B^{\nu\mu}$. The symmetry of the second term in (\ref{Bach2}) is rather obvious. The proof relies on the symmetric properties of the Weyl tensor, which is, by definition the same as that of the Riemann curvature tensor ($C^{\mu\nu\lambda\sigma}=C^{\lambda\sigma\mu\nu}
=-C^{\mu\nu\sigma\lambda}$), as well as of the Ricci tensor ($R_{\lambda\sigma}=R_{\sigma\lambda}$),
\begin{equation}
	C^{\mu\lambda\nu\sigma}R_{\lambda\sigma}
	=C^{\nu\sigma\mu\lambda}R_{\lambda\sigma}
	=R^{\nu\lambda\mu\sigma}R_{\lambda\sigma}.
\end{equation}
Then using the famous relation for the commutation of the second covariant derivatives, one can easily obtain
\begin{equation}
	C^{\mu\sigma\nu\lambda}_{\phantom{\mu\nu\mu\nu};\lambda;\sigma}
	- C^{\mu\sigma\nu\lambda}_{\phantom{\mu\nu\mu\nu};\sigma;\lambda}
	=-(C^{\varkappa\sigma\nu\lambda}R^{\mu}_{\phantom{\nu}\varkappa\lambda\sigma}
	+C^{\mu\sigma\varkappa\lambda}R^{\nu}_{\phantom{\nu}\varkappa\lambda\sigma}).
\end{equation}
Since the Weyl tensor is completely traceless, we can replace $R^{\mu}_{\phantom{\nu}\varkappa\lambda\sigma}$ by $C^{\mu}_{\phantom{\nu}\varkappa\lambda\sigma}$,
\begin{equation}
	C^{\mu\sigma\nu\lambda}_{\phantom{\mu\nu\mu\nu};\lambda;\sigma}
	- C^{\mu\sigma\nu\lambda}_{\phantom{\mu\nu\mu\nu};\sigma;\lambda}
	=-(C^{\varkappa\sigma\nu\lambda}C^{\mu}_{\phantom{\nu}\varkappa\lambda\sigma}
	+C^{\mu\sigma\varkappa\lambda}C^{\nu}_{\phantom{\nu}\varkappa\lambda\sigma}).
\end{equation}
Here the first term on the right-hand-side can be transformed, by making use of the cyclic identity 
($C^{\varkappa\sigma\nu\lambda}=C^{\nu\lambda\varkappa\sigma}=C^{\nu\sigma\lambda\varkappa}-C^{\nu\varkappa\sigma\lambda}$) into
\begin{equation}
	C^{\varkappa\sigma\nu\lambda}C^{\mu}_{\phantom{\nu}\varkappa\lambda\sigma}=
	-C^{\nu\sigma\lambda\varkappa}C^{\mu}_{\phantom{\nu}\varkappa\lambda\sigma}
	+C^{\nu\varkappa\lambda\sigma}C^{\mu}_{\phantom{\nu}\varkappa\lambda\sigma},
\end{equation}
while the second one --- into 
\begin{equation}
	C^{\mu\sigma\varkappa\lambda}C^{\nu}_{\phantom{\nu}\varkappa\lambda\sigma}=
	-C^{\mu\lambda\sigma\varkappa}C^{\nu}_{\phantom{\nu}\varkappa\lambda\sigma}
	-C^{\mu\varkappa\lambda\sigma}C^{\nu}_{\phantom{\nu}\varkappa\lambda\sigma}
	=C^{\mu\sigma\lambda\varkappa}C^{\nu}_{\phantom{\nu}\varkappa\lambda\sigma}
	-C^{\mu\varkappa\lambda\sigma}C^{\nu}_{\phantom{\nu}\varkappa\lambda\sigma}.
\end{equation}
Evidently, their sum is zero. Hence,
\begin{equation}
	B^{\mu\nu}=B^{\nu\mu}.
\end{equation}

Further, we are going to calculate the full covariant derivative of the Bach tensor,
\begin{equation}
	B^{\mu\nu}_{\phantom{\mu\nu};\nu}=
	C^{\mu\sigma\nu\lambda}_{\phantom{\mu\nu\mu\nu};\lambda;\sigma;\nu}
	+\frac{1}{2}C^{\mu\lambda\nu\sigma}_{\phantom{\mu\nu\mu\nu};\nu}R_{\lambda\sigma}
	+\frac{1}{2}C^{\mu\lambda\nu\sigma}R_{\lambda\sigma;\nu}.
\end{equation}
Due to the symmetries we have
\begin{equation}
	C^{\mu\sigma\nu\lambda}_{\phantom{\mu\nu\mu\nu};\lambda;\sigma;\nu}
	=C^{\nu\sigma\mu\lambda}_{\phantom{\mu\nu\mu\nu};\lambda;\sigma;\nu}
	=C^{\mu\lambda\nu\sigma}_{\phantom{\mu\nu\mu\nu};\lambda;\sigma;\nu}
	=\frac{1}{2}(C^{\mu\lambda\nu\sigma}_{\phantom{\mu\nu\mu\nu};\lambda;\sigma;\nu}
	-C^{\mu\lambda\nu\sigma}_{\phantom{\mu\nu\mu\nu};\lambda;\nu;\sigma})
	=\frac{1}{2}(\frac{1}{6}R^{\mu\sigma}R_{;\sigma}-R_{\varkappa\nu;\sigma} 
	R^{\mu\varkappa\sigma\nu}).
\end{equation}
Here we made use of the Bianchi identities and commutation relations. Again, the Bianchi identities and rather lengthy calculations lead us to the conclusion that the Bach tensor is conservative,
\begin{equation}
	B^{\mu\nu}_{\phantom{\nu\nu};\nu}=0.
\end{equation}
Hence, the corresponding part, $T^{\mu\nu}(C^2)$ of the total energy-momentum tensor is also conservative,
\begin{equation}
	T^{\mu\nu}_{\phantom{\nu\nu};\nu}(C^2)=0.
\end{equation}

At last, we investigate the conservative relation for the remaining part,
\begin{equation}
	\mathcal{D}_{\mu\nu}=R_{;\mu\nu}-(R_{;\lambda;\kappa}g^{\lambda\kappa})g_{\mu\nu}-R(R_{\mu\nu}-\frac{1}{4}Rg_{\mu\nu})
\end{equation}
We have
\begin{equation}
	\mathcal{D}^\nu_{\mu;\nu}=
	(R^{;\nu}_{\phantom{\nu};\mu;\nu}
	-R^{;\nu}_{\phantom{\nu};\nu;\mu})
	-R_{;\nu}R^\nu_\mu - RR^\nu_{\mu;\nu} + \frac{1}{2}RR_{;\mu}
\end{equation}
Since $R^\nu_{\mu;\nu}=(1/2)R_{;\mu}$, it follows then,
\begin{equation}
	\mathcal{D}^\nu_{\mu;\nu} = (R^{;\nu}_{\phantom{\nu};\mu;\nu}
	-R^{;\nu}_{\phantom{\nu};\nu;\mu}) - R_{;\nu}R^\nu_\mu\equiv0.
\end{equation}
Here we used the commutation relation. Thus,
\begin{equation}
	\mathcal{D}^{\mu\nu}_{\phantom{\nu};\nu}=0
\end{equation}
and 
\begin{equation}
	T^{\mu\nu}_{\phantom{\nu};\nu}(R^2)=0.
\end{equation}
So, we showed the conservation of the total energy-momentum tensor in Quadratic Gravity, provided that all the tensors and their derivatives, entering the field equations, are well defined, what is true in the bulk.

But, if there exists a jump in the matter distribution, or the $\delta$-function behavior, there appears a singular hypersurface, where the bulk field equations are not valid. In fact, at such a surface two different solutions in two different bulk regions must be linked, using the so-called junction conditions. In General Relativity these are the Israel equations for the thin shells (with $\delta$-like behavior of the energy-momentum tensor), while in Quadratic Gravity they are the equations for the double layers.

Our aim here is to consider the conservative equation on the singular hypersurfaces.  So, let the whole space-time be divided in two different bulk regions, $(+)$ and $(-)$ ones, separating by some singular hypersurface $\Sigma_0$, whose equation is 
\begin{equation}
	n(x^\mu)=0,
\end{equation}
(of course) different in each of the bulk regions. We will be using (as before) the Gauss normal coordinate system associated with $\Sigma_0$,
\begin{equation}
	ds^2=\epsilon dn^2+\gamma_{ij}dx^idx^j, \quad \epsilon=\pm1,
\end{equation}
where the coordinate $n$ runs from $(-)$-region ($n<0$) to $(+)$-region ($n>0$) along the outward normal. And let the total energy-momentum tensor $T^{\mu\nu}$ has the form of the equation (\ref{Tmunu}). The conservative condition takes the form 
\begin{eqnarray}
	T^{\mu\nu}_{\phantom{\mu\nu};\nu}&=&S^{\mu\nu}\delta'(n)n_{,\nu}+S^{\mu\nu}_{\phantom{\mu\nu};\nu}\delta(n)
	\nonumber  \\
	&&+T^{\mu\nu}(+)\delta(n)n_{,\nu} 
	-T^{\mu\nu}(-)\delta(n)n_{,\nu}
	+T^{\mu\nu}_{\phantom{\mu\nu};\nu}(+)\theta(n)
	+T^{\mu\nu}_{\phantom{\mu\nu};\nu}(-)\theta(n).
\end{eqnarray}
Since, as we already know $T^{\mu\nu}_{\phantom{\mu\nu};\nu}(\pm)=0$, then
\begin{equation}
	T^{\mu\nu}_{\phantom{\mu\nu};\nu}=S^{\mu\nu}\delta'(n)n_{,\nu}+S^{\mu\nu}_{\phantom{\mu\nu};\nu}\delta(n)+[T^{\mu\nu}]\delta(n)n_{,\nu}.
\end{equation}
Note that taking the covariant derivative of the $\delta$-function in Gaussian normal coordinates requires some care, as was explained in details in \cite{Senovilla16}. Our article does not address these issues.

According to the rules of the theory of distributions, we must multiply this equation by arbitrary function  $f$ with compact support and integrate  along the coordinate $n$ ($n_{,\nu}=\delta^n_\nu$), the result is 
\begin{equation}
	-(fS^{\mu\nu})_{,n}+f(S^{\mu\nu}_{\phantom{\mu\nu};\nu}+[T^{\mu\nu}])=f(\theta,x^i)C^\mu(x^i).
\end{equation}
Dividing it by $f(0,x^i)$ we get 
\begin{equation}
	b^{(\mu)}(x^i)S^{\mu n}-S^{\mu n}_{\phantom{\mu\nu},n}
	+S^{\mu \nu}_{\phantom{\mu\nu};\nu}
	+[T^{\mu\nu}]=C^\mu(x^i),
\end{equation}
Where $b^{(\mu)}(x^i)=-(f_{,n}/f)(n=0)$ is an arbitrary function of the coordinates on singular hypersurface $\Sigma_0$. In Gauss normal coordinates the above expression is split in $1$ scalar and $1$ vector ($3D$) ones,
\begin{equation}
	\left\{
	\begin{array}{l}	
		(\!b^{(n)}-K)S^{nn}+S^{np}_{\phantom{n}|p} +\epsilon K_{lp}S^{lp} 
		+[T^{nn}]=C^n, \\
		\!(b^{(i)}-K)S^{in}-K^i_lS^{nl}+S^{il}_{\phantom{n}|l} +[T^{in}]=C^i,
	\end{array}	
	\right. 
\end{equation}
where $K_{ij}=-(1/2)\gamma_{ij,n}$ is the extrinsic curvature tensor of hypersurface $\Sigma_0$, and $K=K^l_l$.

Strictly speaking, the above expressions in this very form are valid for Quadratic Gravity, but not for General Relativity, because in the latter case $K_{lp}$ has jumps if $S^{lp}\neq0$. One needs to specify additionally the values of the extrinsic curvature tensor components on $\Sigma_0$. In the next Section~\ref{examples} we will present an example when this is not necessary. Remember, that in General Relativity $S^{nn}=S^{ni}=0$. But, even if there is no thin shell at all, i.\,e., $S^{ij}=0$, we still have a room for the non-conservation of the energy-momentum tensor. It is easy to deduce from the above equations, that the shock waves in the matter distributions, accompanied by the gravitational shock waves (the nonzero jumps in the curvature), are also responsible for the ``matter creation''.

In Quadratic Gravity the jumps provide us  with double layers. We see, that in the absence of the thin shells, it is $S^{nn}$ and $S^{ni}$ that are responsible for the gravitational particle production.

\section{Examples of exact solutions}
\label{examples}

In this section we would like to consider some applications of the theory developed above. Note, that the applications and the theory are quite different things. When constructing a theory, we are allowed to claim: let the whole space-time be divided in two $(\pm)$-regions with the different behavior of the energy-momentum tensor or different vacua, separated by the some singular hypersurface $\Sigma_0$. And let some special conditions be fulfilled at such a hypersurface (in the case of Quadratic Gravity these are the Lichnerowicz conditions). and as the result we obtain some equations on $\Sigma_0$, called the  junction conditions. The situation with the applications, in a sense, reciprocal. We must know the solutions in $(\pm)$-regions. And our aim is just to find such a hypersurface $\Sigma_0$, which satisfied both the the matching equations and some specific conditions (in our case, the Lichnerowicz's ones). And one more important feature: the applications should be as simple as possible in order to make all the machinery quite transparent. For this very reason we have chosen for consideration the spherical symmetric space-times. And in the case of Quadratic Gravity, in addition, we confined ourselves to the conformal gravity, when the linear part is absent (including the cosmological term), and the quadratic parts are combined in the square of the Weyl tensor.

We start with the example from General Relativity and consider a spherically symmetric bubble inside the ``false'' vacuum (described by the de Sitter solution with positive cosmological constant $\Lambda$), and its wall is just the thin shell. The singular hypersurface $\Sigma_0$ is time-like,  the Gauss normal coordinate system line element is 
\begin{equation}
	ds^2=g_{00}(t,n)dt^2-dn^2-r^2(t,n)(d\theta^2+\sin^2\theta d\varphi^2),
\end{equation}
where $r(t,n)$ is the radius of the sphere, the metric on 
$\Sigma_0$ is simply 
\begin{equation}
	d\Sigma^2=d\tau^2-\rho^2(\tau)(d\theta^2+\sin^2\theta d\varphi^2),
\end{equation}
$\rho(\tau)=r(t,0)$, $\tau$ is the proper time of the observer sitting on the shell.

Due to the spherical symmetry, we have only two independent components of the extrinsic curvature tensor, $K^0_0=-(1/2) g_{00.n}|_{\Sigma_0}$ and $K^2_2=K^3_3=-(r_{,n}/r)|_{\Sigma_0}$. Also, we have only two independent components of the surface energy-momentum tensor of the bubble wall, $S^0_0$ and $S^2_2=S^3_3$. The Israel equations are reduced to 
\begin{equation}
	\left\{
	\begin{array}{l}	
		\![K^2_2]=4\pi GS^0_0, \\
		\![K^0_0]+[K^2_2]=8\pi GS^2_2.
	\end{array}	
	\right. 
\end{equation} 

Here we consider the bubble of special kind, first proposed and investigated in \cite{BKT83a,BKT84,BKT87}.

The bubble wall brings no energy, i.\,e., $S^0_0=0$, only surface tension, $S^2_2$ (like bubble in the kettle), it is not empty, the interior is filled with some perfect fluid produced of the released vacuum energy around it (in references cited above, this was called the ``vacuum burning'' phenomenon). The conservative condition equation now becomes
\begin{equation}
	\left\{
	\begin{array}{l}	
		\!-2K^2_2S^2_2+[T^n_n]=C^n, \\ 
		\!-2\frac{\dot\rho}{\rho}S^2_2+[T^{0n}]=C^0.
	\end{array}	
	\right. 
\end{equation} 

Our aim  is demonstrate the very possibility of the matter creation. So, we write down here only the final result without further details:
\begin{equation}
	\left\{
	\begin{array}{l}	
		\!\frac{2}{\rho}\sqrt{\dot\rho^2+1 -\Lambda\rho^2}S^2_2
		-(\Lambda+p)=C^n, \\ 
		\!-2\frac{\dot\rho}{\rho}S^2_2+(\varepsilon+p)\frac{V}{1-V^2}=C^0,
	\end{array}	
	\right. 
\end{equation} 
where $\dot\rho$ is the proper time derivative of the bubble radius $\rho$, $\Lambda$ is the cosmological constant (outside the bubble wall), $\varepsilon$ and $p$ are, respectively, the energy density and pressure of the perfect fluid on the inner side of the bubble wall, and $V$ is the fluid velocity in the inward direction.

The other example comes from the Quadratic Gravity. Again, we assume the spherical symmetry. Moreover, we restrict ourselves to the special case , the conformal gravity, when all the terms in the Lagrangian (\ref{L2}), quadratic in curvature, are just the square of the Weyl tensor, $C^2$. This means, that $\alpha_2=-\alpha_1$ and $\alpha_3=(1/3)\alpha_1$ according to relations $\alpha_2+4\alpha_1=2\alpha_1$ and $\alpha_2+4\alpha_3 =-(2/3)\alpha_1$. The choice stemmed from the fact that we know all spherically symmetric solutions in this theory \cite{Double19,Double20}. The equations for the double layer (only those that are needed for the determination of the singular hypersurface $\Sigma_0$, which now becomes, actually, a world line) are the following (see (\ref{QG1}) and (\ref{QG2})):
\begin{eqnarray}
	\label{QG1b}
	&&\alpha_1\epsilon\,\left(K^{lp}
	-\frac{1}{3}Kg^{lp}\right) [K_{lp,n}]
	= \frac{1}{4}S^{nn},  \\
	\label{QG2b}
	&&\alpha_1\left(g^{il}g^{jp} - \frac{1}{3}g^{ij}g^{lp}\right)[K_{lp,n|j}]
	= \frac{1}{4}S^{in},  
\end{eqnarray}
and we must add the Lichnerowicz conditions 
\begin{equation}
	[K_{lp}]=0.
\end{equation}

The advantage of our choice in favor of the $C^2$ Lagrangian is that the corresponding action is invariant under the conformal transformation. Such a nice property allows us to use the following trick. Writing the spherically symmetric line element in the form 
\begin{equation}
	ds^2=g_{ij}(x)dx^idx^j-r^2(x)(d\theta^2
	+\sin^2\theta d\varphi^2),
\end{equation}
we may choose the radius $r(x)$ as a conformal factor
\begin{equation}
	ds^2=r^2(x)\left(\gamma_{ij}(x)dx^idx^j-(d\theta^2
	+\sin^2\theta d\varphi^2)\right),
\end{equation}
with $\gamma_{ij}=(1/r^2)g_{ij}$, and we can now forget  about the radius, provided $S^{nn}$ and $S^{ni}$ mean now, respectively,  $(1/r^6)S^{nn}$ and $(1/r^6)S^{ni}$ (for details see \cite{Double19,Double20}).

In what follows we will be interested only in time-like double layers and thin shells. So, the conformally transformed line element ($\epsilon=-1$) is
\begin{equation}
	ds^2=\gamma_{00}(\tau,n)d\tau^2-dn^2-(d\theta^2+\sin^2\theta d\varphi^2).
\end{equation}
Thus, we have only one nonzero component of the extrinsic curvature tensor $K_{ij}$, namely
\begin{equation}
	K_{00}=-\frac{1}{2}(\gamma_{00})_{,n}=K,
\end{equation}
where $K$ is the trace of $K_{ij}$, and we put $\gamma_{00}(\tau,0)=1$. Actually, we are dealing now with the $2$-dimensional space-time, and the only invariant quantity that characterizes it is the $2$-dimensional curvature scalar $\tilde R$,
\begin{equation} \label{tildeR}
	\tilde R=\frac{\gamma_{00,nn}}{\gamma_{00}}
	-\frac{1}{2}\left(\frac{\gamma_{00,n}}{\gamma_{00}}\right)^2.
\end{equation}

The total curvature scalar of our $4$-dimensional conformally transformed manifold equals
\begin{equation}
	R=\tilde R-2.
\end{equation}
The equations for determining the trajectory of the double layer become (here a ``dot'' denotes the derivative with respect to the proper time $\tau$):
\begin{equation}
	\left\{
	\begin{array}{l}	
		\!K[\tilde R]=\frac{3}{4\alpha_1}S^{nn}, \\ 
		\![\dot{\tilde R}]=-\frac{3}{4\alpha_1}S^{n0}, \\
		\![K]=0,
	\end{array}	
	\right. 
\end{equation} 
and the conservative relation now reads
\begin{equation}
	\left\{
	\begin{array}{l}	
		(b^{(n)}-K)S^{nn}+\dot S^{n0} +\epsilon KS^{00}[T^{nn}]=C^n \\
		(b^{(0)}-2K)S^{0n}+\dot S^{n0}+[T^{0n}]=C^0.
	\end{array}	
	\right. 
\end{equation} 
Since we decided to consider only time-like singular hypersurfaces, the function $C^0(\tau)$ describes the energy density creation, while $C^n(\tau)$ is responsible for the creation of the energy density flow.

Everybody knows that in General Relativity all spherically symmetric vacuum solutions belong to the two-parametric family. The parameters are the Schwarzschild mass $m$ and the cosmological constant $\Lambda$. In conformal gravity the spherically symmetric vacuum solutions belong to one of three classes. They are:

{\bf Class I}:

\begin{equation}
	\tilde R=2, \quad C^2=0.
\end{equation} 
All the cosmological models, i.\,e., homogeneous and isotropic manifolds enter this class, including the flat Minkowski space-time and de Sitter ($m=0$, $\Lambda>0$) and anti-de Sitter ($m=0$, $\Lambda<0$) ones.

{\bf Class II}:

\begin{equation}
	\tilde R=-2, \quad C^2=\frac{16}{3}.
\end{equation} 
The solutions  of this class are obtained from that ones of Class I by interchanging the temporal and spatial variables.

{\bf Class III}:

This is the one-parametric family. The corresponding two-dimensional conformally transformed line element is
\begin{equation} \label{ClassIII}
	ds_2^2=Adt^2-\frac{d\tilde R^2}{A}, \quad 
	A=\frac{1}{6}(\tilde R^3-12\tilde R+C_0), \quad C_0=const.
\end{equation} 
The spherically symmetric vacuum solutions of General Relativity with $\Lambda\neq0$ belong to this class, in this case $C_0=16-2(12Gm)^2\Lambda$.

Our aim is to consider different combinations in $(\pm)$-regions and find the corresponding trajectories of the double layers in-between. We begin with (I--II)-case, i.\,e., on one side of the singular hypersurface $\Sigma_0$ one has a solution from Class~I, while on the other side --- from Class~II. Note that the Classes have to be different, otherwise there will be no double layer at all. Thus,
\begin{equation}
	[\tilde R]=\pm4=const, \quad \dot{\tilde R}=0,
\end{equation} 
and we are left with the equation
\begin{equation}
	K=\pm\frac{3}{16\alpha_1}=S^{nn}.
\end{equation} 
But, we also have the equation (\ref{tildeR}), relating $\gamma_{00,n}$ and $\tilde R$. For $\tilde R=\pm2$ one gets 
\begin{equation} \label{tildeRb}
	\pm2=\frac{\gamma_{00,nn}}{\gamma_{00}}
	-\frac{1}{2}\left(\frac{\gamma_{00,n}}{\gamma_{00}}\right)^2.
\end{equation}
This equation can easily be solved for $Z=-(1/2)(\gamma_{00,n}/\gamma_{00})$. The result is
\begin{eqnarray} \label{tildeRb}
	Z&=&-\tanh\left(n+f_{(+)}(\tau)\right) \quad \mbox{for} 
	\quad \tilde R=+2, \\
	Z&=&\tan\left(n+f_{(-)}(\tau)\right) \quad \mbox{for} 
	\quad \tilde R=-2,
\end{eqnarray}
where $f_{(\pm)}(\tau)$ are arbitrary real-valued functions. On $\Sigma_0$, where $n=0$,
\begin{equation} 
	Z|_{\Sigma_0}=K.
\end{equation}
and the only way to obey the Lichnerowicz conditions is to put $f_{(+)}(\tau)=f_{(-)}(\tau)=0$. This means that
\begin{equation} 
	K=0.
\end{equation}
Thus we have got the static solution (on both sides of $\Sigma_0$ the curvature $\tilde R$ is constant) without matter creation ($S^{nn}=S^{ni}=0$). Both vacua remain to be vacua, and no collapse of the double layer. Note, that, despite of the zero values for $S^{nn}$ and $S^{n0}$ we can still have $\dot S^{00}=C^0$ ($C^n=0$, as it should be, because we postulated the vacuum states in $(\pm)$-regions). Such a situation can be called the ``emergent thin shell''. And the source of this process is hidden  inside the thin shell itself.

Let us consider the case when  a double layer matches some solution from Class~I (or Class~II) on the one side (with $\tilde R=\pm2$) and some solution from Class~III on the other side. The requirement that both vacua remained empty dictates $S^{n0}=0$, hence $[\dot{\tilde R}]=0$, and the equation  for the singular hypersurface on that side where we have the solution from the Class~III is simply $\tilde R=\tilde R_0=const$. In this region the two-dimensional line element has the form from equation (\ref{ClassIII}). In order the double layer be time-like, it is necessary that $A(\tilde R)>0$. The normal coordinate to the surface $\tilde R=const$ is collinear  with our coordinate $\tilde R$, i.\,e., 
\begin{equation} 
	dn=\frac{d\tilde R}{\sqrt{A}}, \quad \tilde R_{,n}=\sqrt{A}.
\end{equation}
Further,
\begin{equation} 
	\gamma_{00}=\frac{A(\tilde R)}{A(\tilde R_0)}, \quad 
	K=-\frac{1}{2}\gamma_{00,n}(\tilde R_0)
	=-\frac{\tilde R_0^2-4}{4\sqrt{A(\tilde R_0)}}=K_0.
\end{equation}
The Lichnerowicz conditions can be easily satisfied by the appropriate choice of $f_{(\pm)}(\tau)$ for $\tilde R=\pm2$ vacua. We see that $S^{nn}$ is constant and, in general, nonzero. The substitution of all these into the conservative relations shows that they can be always satisfied by the appropriate choice of the ``arbitrary'' function $B^{(n)}$ (remember, that $[T^{nn}]=[T^{0n}]=C^n = 0$ in the case of the vacuum solutions in the $(\pm)$-regions). Again, we may have a situation of the ``emergent thin shell'', but this time the source is the double layer ($S^{nn}\neq0$). 

At last, consider the case when on both sides of the singular hypersurface $\Sigma_0$ the vacuum solutions belong to Class~III. All the formulas were already written above. Surely, choosing the appropriate constants of integration, it is always possible to make the normal coordinate $n$ continuous across $\Sigma_0$. Then, the requirement for the solutions in the bulk to be the vacuum ones translates into $[\tilde R]=const$, and together with the Lichnerowicz condition, the latter leads to constant values for $\tilde R$ on both sides of $\Sigma_0$. Therefore, again we have the static solution --- no collapse. 

The overall result is that in conformal gravity there are no collapsing spherically symmetric double layers without radiation! 

Suppose now, that there are no double layers --- only thin shells. Then, as we already know, $S^{nn}=0$,  $S^{ni}=0$, and the analog of the Israel equation for the spherically symmetric conformal gravity takes the form 
\begin{eqnarray} 
	[K_{,nn}]&=&-\frac{3}{8\alpha_1}S^{00}=-\frac{3}{8\alpha_1}S^0_0,
	\\
	{}[K_{,nn}]&=&-\frac{3}{4\alpha_1}S^{22}=\frac{3}{4\alpha_1}S^2_2,
\end{eqnarray}
Note, that $TrS_{ij}=S^0_0+2S^2_2=0$. To these equations one must add the Lichnerowicz condition ($[K_{ij}]=0$) and conditions for the double layer absence ($[K_{lp,n}]=0$). In our case they look as follows
\begin{equation} 
	[K]=0, \quad [K_{,n}]=0.
\end{equation}
This means that 
\begin{equation} 
	[\tilde R]=0.
\end{equation}
Since on both sides of the time-like $\Sigma_0$ we have different vacuum solutions, the corresponding conservative relations are now very simple
\begin{equation}
	\left\{
	\begin{array}{l}	
		\!-KS^0_0=C^n, \\ 
		\!\dot S^0_0=C^0.
	\end{array}	
	\right. 
\end{equation} 

Consider now different combinations of the vacuum solutions in $(\pm)$-regions. Let, first, on both sides the solutions belong to Class~I (or Class~II, they must be the same). We have already introduced above the function $Z=-(1/2)(\gamma_{00,n}/\gamma_{00})$, which equals $-\tanh\left(n+f_{(+)}(\tau)\right)$ for $\tilde R=+2$ and $Z=\tan\left(n+f_{(-)}(\tau)\right)$ for $\tilde R=-2$. It easy to see that from the Lichnerowicz condition $[K]=0$ it follows both $[K_{,n}]=0$ and $[K_{,nn}]=0$. Hence $S^0_0=S^2_2=0$, what means that for such a combination the thin shells without the double layers do  not exist at all. 

Now let on the one side of $\Sigma_0$ the vacuum solution is from Class~I (or Class~II), while on the other side --- from Class~III. The result (without details) is the following: due to the continuity of the curvature at $\Sigma_0$, we may have only static solutions with $\tilde R=2$ (or $\tilde R=-2$), but they exist only for the specific values of the parameters $C^0$ in the metric from Class~III. So, no collapse.

The last of all possible combinations of vacuum solutions on $(\pm)$-regions is the case when both of them belong to Class~III. This situation is most difficult for investigation, the main problem being to satisfy the Lichnerowicz condition. But the result is rather nice and simple. Namely, the collapse is possible, all the trajectories of the thin shell, $\tilde R(\tau$), obey the universal second order differential equation
\begin{equation}
	\ddot {\tilde R}=1-\frac{1}{4}\tilde R^2,
\end{equation} 
which can be easily solved in quadrature,  and the analogue of the Israel equations is 
\begin{equation} \label{ClassIIIb}
	\sigma_{(+)}\sqrt{C_0(+)+V_0^2}-\sigma_{(-)}\sqrt{C_0(-)+V_0^2}
	=\frac{3}{4\alpha_1}S^0_0,
\end{equation} 
where $C_0(\pm)$ are the values of the only parameter $C_0$ of the solutions in $(\pm)$-regions, $V_0=\dot {\tilde R}(\tau=0)$ is the initial ``velocity'' of the curvature,  and $\sigma=\pm1$ depending on which of the sides of the solutions ($\tilde R=\pm\infty$) ``looks'' at matching hypersurface $\Sigma_0$. Therefore $S^0_0=const$. Moreover, the dynamical equation above guaranties the $K=0$ on both sides of $\Sigma_0$. Thus, there is no creation of matter and no radiation coming from the thin shell. See details for derivation of equation (\ref{ClassIIIb}) in Appendix~\ref{appendix}.

\section{Conclusions and discussions}

In this concluding Section we would like to summarize the obtained results. We have chosen for our investigation the Quadratic Gravity and its connections with the conservation of the energy-momentum tensor for the matter fields. The choice was motivated by the fact that it is the quadratic combinations of the Riemann curvature tensor and its contractions that appear in the trace anomalies in th one-loop approximation of the quantum field theories on the curved space-time background. 

First of all, we rearranged the terms on the Lagrangian of the Quadratic Gravity and, instead of the commonly used ``naked'' squared curvatures, we wrote down it as the sum of the Weyl tensor, the Gauss-Bonnet term and the square of the curvature scalar. Then we proved explicitly, that the right-had (gravitational) side of the field equations is conservative, and so does the total energy-momentum tensor. Of course, for the linear part of the total Lagrangian it was the well known fact for the whole century, as well as for the $C^2$-part, but we presented the proof for the sake of completeness.

These conservative conditions are valid, provided the Riemann curvature tensor and its contractions (Ricci tensor and curvature scalar) are well-defined, i.\,e., in the bulk. But, both the General Relativity and Quadratic Gravity provide us with the possibility of the existence of the so-called singular hypersurfaces.

In General Relativity the singular hypersurface $\Sigma_0$ appears only when the matter energy-momentum tensor contains the $\delta$-function part, it is called the thin shell. The corresponding Einstein equations on these shells are the Israel equations which relate the surface energy-momentum tensor, $S_{ij}$ ($i,j$ are the coordinate indices on $\Sigma_0$), to the jumps in the extrinsic curvature tensor, $K_{ij}$, describing the embedding of the three-dimensional singular hypersurface into the four-dimensional space-time.

In Quadratic Gravity the situation is much more interesting. First, we are not allowed to have $\delta$-function behavior in the curvature (unlike in General Relativity), since, in generic case, this would lead to the appearance of $\delta^2$-term in the Lagrangian what is forbidden by the conventional theory of distributions. So, we are obliged to impose the Lichnerowicz conditions $[K_{ij}]=0$. Thus, the Riemann curvature tensor and its contractions may have at most jumps across singular hypersurface $\Sigma_0$, they appear in the form  $[K_{lp,n}]\neq0$ ($n$ is the coordinate, normal to $\Sigma_0$). Second, since the field equations in Quadratic Gravity are of the fourth order in derivatives of the metric tensor (of second order in derivatives of the curvature), these jumps lead not only to the $\delta$-terms, but also to the $\delta'$-terms, the former being describe the thin shells, while the latter --- the so-called double layer, discovered recently by J.M.M.~Senovilla. Double layers can be considered as the gravitational shock waves. Following the rules of the theory of distributions, one must to integrate the $\delta$- and $\delta'$-terms, together with the arbitrary function of four variables having a compact support. The result is the appearance of the arbitrary functions of three variables in the field equations on $\Sigma_0$. These functions should be then determined by solving matching equations for every specific choice of the solutions in the bulk and surface energy-momentum tensor $S_{ij}$ of the corresponding thin shell. Our approach to deriving the field equations on the singular hypersurface, described shortly in the present paper, is based on the least action principle. The $\delta'$-function is not even mentioned, but the allusion of its existence is there, taken the form of the arbitrary tensorial function connecting the variations of the extrinsic curvature tensor, $\delta K_{ij}$, with that of the metric tensor, $\delta \gamma_{ij}$. Third, the appearance of, possibly, nonzero components, $S^{nn}$ and $S^{ni}$, of the surface energy-momentum tensor. J.M.M.~Senovilla, who discovered this phenomenon and emphasize its necessity, called them the external pressure and external flow, correspondingly. They are rather unusual things, because it is only $S^{ij}$ ($S_{ij}$) that describe the energy content of the thin shells. Note, that in General Relativity $S^{nn}=0$ and $S^{ni}=0$ by virtue of the field equations.

We derived the conservative conditions in the case when the energy-momentum tensor on the singular hypersurface contains both a jump and a $\delta$-function term. Since the existence of the $\delta'$-term in the resulting equation, its subsequent integration leads to the appearance of the ``arbitrary'' functions which, quite evidently, are connected with that ``arbitrary'' functions (and their normal derivatives) entering the field equations for the double layers. It is seen at ones, that, in the absence of the thin shell ($S^{ij}=0$) it is the ``unusual'' $S^{nn}$ and $S^{ni}$, that are responsible for the possible non-conservation of the energy-momentum tensor on the singular hypersurface $\Sigma_0$. Thus, the nature of these components becomes clear: they describe, phenomenologically, the creation of matter by the geometry ``inside'' the gravitational shock waves.  

In the remaining part of the paper we considered some applications of the obtained theoretical results. Our choice is the spherically symmetric conformal gravity, what was dictated by our knowledge of all vacuum solutions in this case \cite{bde1,bde2}. The main result is that the time evolution of the singular hypersurface with the double layer between two different vacua, i.\,e., its collapse, is impossible without radiation.

\appendix*
\section*{Appendix A}
\label{appendix}

In this Appendix we present the details of derivation of the equation (\ref{ClassIIIb}) for the trajectories of the time-like thin shells in the  spherically symmetric conformal gravity (see the end part of Section~\ref{examples}), when double layer is absent, and on both sides of the the thin shell the vacuum  solutions belongs to Class~III. The latter means that the two-dimensional line element (after removing the line element of the unit sphere) has the form (\ref{ClassIII}), where $\tilde R$ is the two-dimensional curvature scalar, and $C_0$ is the only parameter of the Class~III vacuum solutions.

Let the equation of the thin shell trajectory be $\tilde R=R_0(\tau)$, and $n$ be a coordinate in the direction of outward normal vector to the trajectory, running from $(-)$-region to $(+)$-region, with $n=0$ on it. Then, the line element can be written in the form 
\begin{equation}
	ds^2_2=\gamma_{00}(\tau,n) -dn^2, \quad \gamma_{00}(\tau,o)=1.
\end{equation} 
The only component of the extrinsic curvature tensor is
\begin{equation}
	K_{00}=-\frac{1}{2}\gamma_{00,n}, \quad
	K=-\frac{1}{2}\gamma^{00}\gamma_{00,n}
	=-\frac{1}{2}\frac{\gamma_{00,n}}{\gamma_{00}}.
\end{equation} 

The equations governed the thin shell trajectory are jumps of the trace $K$ and its normal derivative,
\begin{equation}
	[K]=0, \quad [K_{,n}]=0.
\end{equation} 
The first one is the Lichnerowicz condition, unavoidable in Quadratic Gravity, while the second is the condition for the absence of the double layer. The analogue of the Israel equations for thin shells is
\begin{equation}
	[K_{,nn}]=-\frac{3}{8\alpha_1}S^{00},
\end{equation} 
where $S^{00}$ is the surface energy density of the shell.

The two-dimensional curvature scalar $\tilde R$ equals
\begin{equation}
	\tilde R=\left(\frac{\gamma_{00,n}}{\gamma_{00}}\right)_{,n}
	-\frac{1}{2}\left(\frac{\gamma_{00,n}}{\gamma_{00}}\right)^2
	=-2(K_{,n}+K^2).
\end{equation} 
From the conditions, imposed above,  
\begin{equation}
	[\tilde R]=0
\end{equation} 
and vice verse, if we impose the condition $[\tilde R]=0$ and satisfy the Lichnerowicz condition $[K]=0$, then the absence of the double layer will be guaranteed.

Let us make the transformation from the coordinates $(t,\tilde R)$ to $(t,n)$ in $(\pm)$-regions (equation (\ref{ClassIII})),
\begin{eqnarray}
	ds^2_2&=&Ad\eta^2-\frac{1}{A}d\tilde R^2
	=\gamma_{00}d\tau^2-dn^2, \nonumber \\
	t&=&t(\tau,n), \quad \dot t=t_{,\tau} 
	, \quad  t_n=t_{,n}, \nonumber \\
	\tilde R&=&R(\tau,n), \quad \dot{\tilde R}=\tilde R_{,\tau} 
	, \quad  \tilde R_n=\tilde R_{,n},
\end{eqnarray} 
then,
\begin{equation}
	\left\{
	\begin{array}{l}	
		\!A\dot t^2-\frac{1}{2}\dot{\tilde R}=\gamma_{00}, \\ 
		\!A\dot t t_{,n}-\frac{1}{A}\dot{\tilde R}\tilde R_n=0, \\ 
		\!\frac{1}{A}\tilde R_n^2 -At^2_n=1.
	\end{array}	
	\right. 
\end{equation}
It follows from this, that
\begin{equation}
	\gamma^{00}\dot{\tilde R}^2-\tilde R^2_n=A, \quad
	\tilde R=\pm\sqrt{\gamma^{00}\dot{\tilde R}^2-A}
	=\sigma\sqrt{\gamma^{00}\dot{\tilde R}^2-A},
\end{equation}
where $\sigma=\pm1$, indicating whether $\tilde R$is increasing in the $n$-direction or decreasing.

Let us rewrite the transformation relation in the form 
\begin{equation}
	\left\{
	\begin{array}{l}	
\!t_n^2=\frac{1}{A}\left(\frac{1}{A}\tilde R_n^2-1\right) =M^2, \\ 
\!\dot t^2=\frac{1}{A}\left(\gamma_{00} + \frac{1}{A}\dot{\tilde R}^2\right)=L^2,  \\ 
\!LM=\frac{1}{A^2}\dot{\tilde R}\tilde R_n.
	\end{array}	
	\right. 
\end{equation}
One must add also the integrability condition 
\begin{equation}
	L_{,n}=\dot M.
\end{equation}
Extracting, then, $\gamma_{00}$,
\begin{equation}
	\gamma_{00}=AL^2-\frac{\dot{\tilde R}}{A},
\end{equation}
we managed to calculate  $\gamma_{00,n}$ on the shell (from either sides). The calculations are cumbersome, but the result is surprisingly simple,
\begin{equation}
	\gamma_{00,n}=\frac{2\ddot R_0+A^{'}(R_0)}{\tilde R_{,n}}
	=-2K.
\end{equation}
From Lichnerowicz condition it then follows the universal dynamical equation for the thin shells, 
\begin{equation} \label{dynamical}
	\ddot R_0=1-\frac{1}{4}R_0^2,
\end{equation}
and, simultaneously, that (on both sides)
\begin{equation}
	K=0.
\end{equation}
The first integral for the dynamical equation (\ref{dynamical}) can be easily found, 
\begin{equation}
	\dot{\tilde R}_0^2=\frac{1}{6}(12R_0-R_0^3)+V_0^2, \quad \rightarrow \quad
	\tilde R_n^2=\dot{\tilde R}_0^2+A=C_0+V_0^2,
\end{equation}
and, noticing, that $[K_{,n}]=-(1/2)[\tilde R_n]$, we get the requested equation (\ref{ClassIIIb}):
\begin{equation} 
	\sigma_{(+)}\sqrt{C_0(+)+V_0^2}-\sigma_{(-)}\sqrt{C_0(-)+V_0^2}
	=\frac{3}{4\alpha_1}S^0_0,
\end{equation} 
Consequently,
\begin{equation}
	S^0_0=const.
\end{equation}

\begin{acknowledgements}
We are grateful to E. O. Babichev for stimulating discussions. This research was supported in part by the Russian Foundation for Basic Research project n. 18-52-15001-NCNIa. 
\end{acknowledgements}

\end{document}